# A Synergistic Framework Leveraging Autoencoders and Generative Adversarial Networks for the Synthesis of Computational Fluid Dynamics Results in Aerofoil Aerodynamics

[1]Tanishk Nandal [2]Vaibhav Fulara, [3]Raj Kumar Singh
[1] Delhi Technological University ,[2] Delhi Technological University [3] Delhi Technological University
[1] tanishknandal_2k19me252@dtu.ac.in,[2] vaibhavfulara_2k19me262@dtu.ac.in, [3]rajkumarsingh@dtu.ac.in

*Abstract—* In the realm of computational fluid dynamics (CFD), accurate prediction of aerodynamic behaviour plays a pivotal role in aerofoil design and optimization. This study proposes a novel approach that synergistically combines autoencoders and Generative Adversarial Networks (GANs) for the purpose of generating CFD results. Our innovative framework harnesses the intrinsic capabilities of autoencoders to encode aerofoil geometries into a compressed and informative 20-length vector representation. Subsequently, a conditional GAN network adeptly translates this vector into precise pressure-distribution plots, accounting for fixed wind velocity, angle of attack, and turbulence level specifications. The training process utilizes a meticulously curated dataset acquired from JavaFoil software, encompassing a comprehensive range of aerofoil geometries. The proposed approach exhibits profound potential in reducing the time and costs associated with aerodynamic prediction, enabling efficient evaluation of aerofoil performance. The findings contribute to the advancement of computational techniques in fluid dynamics and pave the way for enhanced design and optimization processes in aerodynamics.

## I. INTRODUCTION

In aerospace engineering, Computational Fluid Dynamics (CFD) plays a paramount role in predicting the aerodynamic behavior of an aircraft. Traditional CFD approaches involve using wind tunnels and simulation tools to test the aerodynamics of various aerofoils. However, these methods are time-consuming and expensive, making it challenging to perform tests on a large number of aerofoils. Hence, there is a need for faster and more efficient methods to predict aerodynamic performance.

Deep Learning (DL) has recently shown tremendous potential in various applications, including CFD. DL algorithms can learn from a large dataset of aerodynamic simulations to predict the aerodynamic behavior of an aerofoil for different wind velocities, angles of attack, and turbulence levels. However, the prediction accuracies are totally dependent on the quality of training data.

To address this problem, we propose a novel approach that combines an autoencoder and a Generative Adversarial Network (GAN) for the generation of CFD results for a given aerofoil geometry. Our GAN model takes a 20-length vector as input, generated by an autoencoder, designed primarily to encode the aerofoil geometry in a compressed form. The GAN then converts the input vector into an output result pressure-distribution plot for a fixed wind velocity, angle of attack, and turbulence level.

The training data for our model has been acquired from JavaFoil software, which simulates the flow over an aerofoil section using a viscous-inviscid interaction technique. We trained our model on a dataset of 4096 simulations, each with a different aerofoil geometry, out of which 3276 simulation samples are used for training, and the remaining 820 samples are used for validation.

Our model has the potential to significantly reduce the time and cost involved in predicting the aerodynamic performance of an aerofoil. While the traditional CFD approaches can be reserved for later screening of the shortlisted aerofoils, our model can be used for initial shortlisting, thereby reducing the overall testing time and cost. The approach can be extended to other areas of aerospace engineering, such as wing design, propeller design, and flight control systems, to name a few.

Our work highlights the potential of combining ML with CFD to predict the aerodynamic behavior of an aerofoil. This approach provides a faster and more efficient method of predicting aerodynamic performance, which could significantly impact the aerospace industry.

In recent years, the application of machine learning in CFD has garnered significant attention due to its potential to enhance the simulations. Researchers have explored various machine learning techniques to accelerate simulations, reduce computational costs, and improve the accuracy of fluid flow predictions. Ling et al. [1] proposed a deep neural network as an alternative to traditional turbulence models in CFD simulations. By training the network using a Reynolds-averaged Navier-Stokes (RANS) simulations dataset, they achieved notable advancements in turbulence prediction, surpassing the capabilities of conventional RANS models. Kutz et al. [2] introduced machine learning techniques to model fluid flows within a low-dimensional state space. Their method involved combining dynamic mode decomposition with sparse regression to identify the governing equations of fluid dynamics. This approach exhibited promising results in accurately predicting flow behavior and showed potential for reducing the computational burden associated with CFD simulations.

Building upon these ideas, Wang et al. [3] proposed a hybrid approach that combined machine learning algorithms with numerical simulations to model turbulent flow over aerofoils. By leveraging a convolutional neural network, they successfully learned the intricate features of the flow field and improved the accuracy of numerical simulations. This integration of machine learning and traditional simulation methods showcased enhanced performance in capturing the complex phenomena associated with turbulent flows.

Another study by Wang et al. [4] focused on leveraging machine learning algorithms to optimize the design of microfluidic mixers. Their approach involved employing a genetic algorithm to generate initial design parameters and training a neural network to predict mixer performance. The results indicated a significant improvement over the traditional trial-and-error approach, highlighting the potential of machine learning in enhancing the efficiency and effectiveness of fluid system design.

Exploring alternative avenues, researchers have also delved into generative adversarial networks (GANs) to generate CFD results. Lapeyre et al. [5] proposed employing GANs to generate realistic turbulent flow fields by conditioning the GAN on input parameters such as Reynolds number and pressure gradient. This conditional GAN framework demonstrated the capability to generate highly accurate and realistic flow fields, offering promising possibilities for various CFD applications. Similarly, Jin et al. [6] applied GANs to create the pressure distribution on aerofoils, focusing on parameters such as angle of attack and Mach number. The GAN-based approach successfully generated precise pressure distributions across a wide range of input parameters, highlighting its potential for optimizing aerodynamic design and analysis.

In addition to turbulence modeling and flow field generation, researchers have explored the application of machine learning algorithms for improving the accuracy of turbulence models. Wang et al. [7] proposed a machine learning algorithm to correct errors in RANS turbulence models. By training a neural network to learn the required corrections, they achieved improved accuracy and better agreement with experimental data.

Expanding the scope of machine learning in CFD, Smith et al. [8] investigated machine learning for unmanned aerial vehicle (UAV) aerodynamic performance prediction. Their approach showcased the potential of machine learning in optimizing UAV design and enhancing flight efficiency. Zhang et al. [9] explored the use of deep learning for flow pattern recognition in multiphase flows. Training deep neural networks on large datasets achieved accurate flow pattern identification, contributing to a better understanding and analysis of multiphase flow phenomena.

Chen et al. [10] applied recurrent neural networks (RNNs) for improved turbulence modeling. The sequential nature of RNNs allowed them to capture temporal dependencies in flow data, resulting in enhanced turbulence predictions. Liu et al. [11] focused on combining machine learning and data assimilation techniques for weather forecasting. By assimilating observational data into a machine learning-based weather model, they achieved improved forecasting accuracy, which has important implications for weather-dependent industries and emergency preparedness.

Guo et al. [12] investigated the application of deep reinforcement learning for aerodynamic shape optimization. Their work demonstrated the potential of using reinforcement learning algorithms to autonomously optimize aerodynamic shapes, leading to enhanced performance and efficiency in aerospace engineering.

In another direction, Li et al. [13] explored the use of transfer learning in CFD simulations. By transferring knowledge from pre-trained models to new domains or tasks, they achieved accelerated simulations without sacrificing accuracy, offering a promising avenue for reducing computational costs in CFD. Also, Wang et al. [14] proposed a hybrid approach that integrated physics-based models with machine learning techniques for multiphase flow simulations. By combining the strengths of both systems, they demonstrated improved accuracy and efficiency in simulating complex multiphase flows.

These studies collectively emphasize the significant potential of machine learning in various aspects of CFD, ranging from turbulence modeling and flow pattern recognition to optimization, weather forecasting, and simulation efficiency improvements. By integrating machine learning techniques, researchers continue to pave the way for advancements in CFD, fostering a deeper understanding of fluid dynamics and enabling more efficient and accurate simulations.

## II. METHODOLOGY

### A. Autoencoders

Autoencoders are neural networks expected to learn a compressed input data representation. These networks chiefly consist of two main parts: an encoder and a decoder. The encoder is fed the input data and produces a compressed representation using the input, often referred to as a "latent code". The decoder then uses this to produce an output that tries to reconstruct the original input data.

Autoencoders are often used for dimensionality reduction, where the high-dimensional input data is reduced to a lower-dimensional representation that still preserves the critical features of the input data. They can also be used for data compression and denoising. The input data is corrupted with noise, and the autoencoder learns to remove the noise to reconstruct the original data.

In recent years, autoencoders have become an essential tool in ML and have been applied to a wide range of applications, such as anomaly detection [15], natural language processing, and image and speech recognition [16]-[17]. Variants of autoencoders, such as denoising [18] and variational autoencoders [19], have also been developed to improve their performance and applicability to various tasks.

### B. Dataset

JavaFoil uses a panel approach to discretize the aerofoil into a sequence of panels and determine the flow field at each panel to solve the potential flow equations. The flow around the aerofoil is computed by superimposing the effects of each panel, which is viewed as a source of possible flow.

Additionally, the approach meets the no-slip boundary requirement at the aerofoil surface. After calculating the flow field, JavaFoil applies the Bernoulli equation to determine the aerofoil surface's pressure coefficient at each location. We sample 4096 aerofoils based on NACA 4-digit profiles by changing the aerofoil geometry parameters to train the proposed model. The thickness $t/c$ ranges from 0.1 to 0.18, with the Camber $f/c$ ranging between 0 to 0.08. The Camber location $x_f/c$ is varied from 0.4 to 0.6. Pressure coefficient $C_p$ field around the aerofoil for the 4096 samples is simulated using Javafoil. The raw geometry image and the CFD results are then used for training.

*C. Convolutional Autoencoder*

Let $x$ represent a vector of input data in the high-dimensional space X. The encoder function $f(x)$ converts the input value $x$ to a latent representation $z$ with a lower dimension, learning a mapping $f: x \to z$. The decoder function $g(z)$ maps $z$ back to a reconstruction $x'$ in the original high-dimensional space X, mapping $g: z \to x'$. The autoencoder is trained to minimise the reconstruction error, which is commonly expressed as the mean squared error, between the original data $x$ and the reconstructed data $x'$. A simple representation of an autoencoder is shown in Fig. 1.

The core operation of a convolutional layer is the convolution operation. It involves sliding a filter over the input data and computing the dot product between the filter and the corresponding input region at each position. The kernel acts as a window or template that extracts certain features or characteristics from the input. Given an input image and a kernel, this process generates a feature map, which represents the presence of specific elements or patterns in the input data, given by:

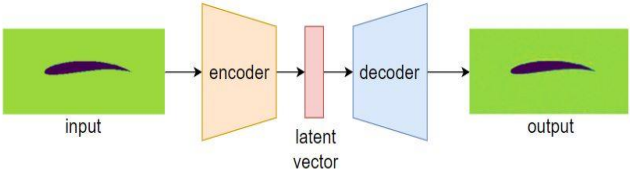

**Fig 1:** A simple representation of the Autoencoder network

$$F(i,j) = \sum_x \sum_y I(x,y) * K(i-x, j-y) \quad (1)$$

In our model, the autoencoder is designed to take in a matrix of size $164 \times 364 \times 1$ which is a binary representation of the input geometry. The autoencoder is trained to learn a representation of this geometry as a latent vector of dimension 20. It is then reconstructed back to the original dimensions of $164 \times 364 \times 1$. The encoder part of the network consists of two 2D convolutional layers with kernel size 3, each followed by a max pooling and a ReLU activation. It is then flattened and passed into a linear layer that forms the latent layer.

The decoder part of the network does the opposite; it attempts to reconstruct the original image using the latent vector obtained at the end of the encoder network. It consists of three 2D convolution layers and ReLU activation, two of which are preceded by upsampling layers that scale their inputs by a factor of 2 using nearest neighbor interpolation.

Mean squared error (MSE) is used as the loss function for the autoencoder, which minimizes the average squares of differences between the original image and its reconstruction:

$$MSE = \frac{1}{n} \sum_1^n (p_i - p'_i)^2 \quad (2)$$

where $p$ represents pixels of the original image and $p'$ represents pixels of the reconstructed image, and $n = h \times w \times c$ is the total number of pixels for an image of height $h$, width $w$, having $c$ channels. The detailed autoencoder architecture is shown in Fig. 2.

*D. Generative Adversarial Networks*

Generative Adversarial Networks (GANs) are deep-learning neural networks capable of generating synthetic data similar to real-world data. These image generators are composed of a generator and a discriminator. The former takes in a random noise vector as input and generates fake data samples, while the latter tries to distinguish between the developed models and actual samples. The generator attempts to trick the discriminator into accepting the generated samples as correct while these two components are being trained together in a minimax game. The discriminator, on the other hand, aims to categorize the real and fake samples appropriately.

The key advantage of a GAN is that it can generate data samples that are indistinguishable from actual data, which can be helpfully used for various purposes, including, but not limited to, image generation, video generation, and even data augmentation. GANs have also been effective in unsupervised learning tasks, where the network is trained on unlabeled data and can learn valuable data representations. However, GANs are challenging to train and can suffer from problems similar to mode collapse, wherein the generator can produce just a limited set of outputs, and vanishing gradients, where the gradients in the training process become too small to be useful. Despite these challenges, GANs have proven to be a powerful tool in the field of deep learning and continue to be an active area of research.

Given random noise $z$, and real data $x$, the vanilla GAN objective can be formulated as follows:

$$min_G max_D \mathcal{L}_{vGAN}(G,D)$$
$$= \mathbb{E}_x[log D(x)] + \mathbb{E}_z\left[\log\left(1 - D\big(G(z)\big)\right)\right] \quad (3)$$

The Generator $G$ tries to minimize this objective while the Discriminator $D$ tries to maximise it, leading to a two player minimax game.

Conditional Generative Adversarial Networks (cGANs) are an extension of GANs that allow for the conditional generation of data. The critical distinction between GANs and cGANs is that cGANs incorporate additional conditioning information, such as class labels or other structured metadata, to produce more precise and regulated outputs for both the generator and discriminator. cGANs may generate data samples that adhere to specific qualities or characteristics by conditioning the generator on this information, making them especially helpful for applications like image and video synthesis, text-to-image translation, and style transfer.

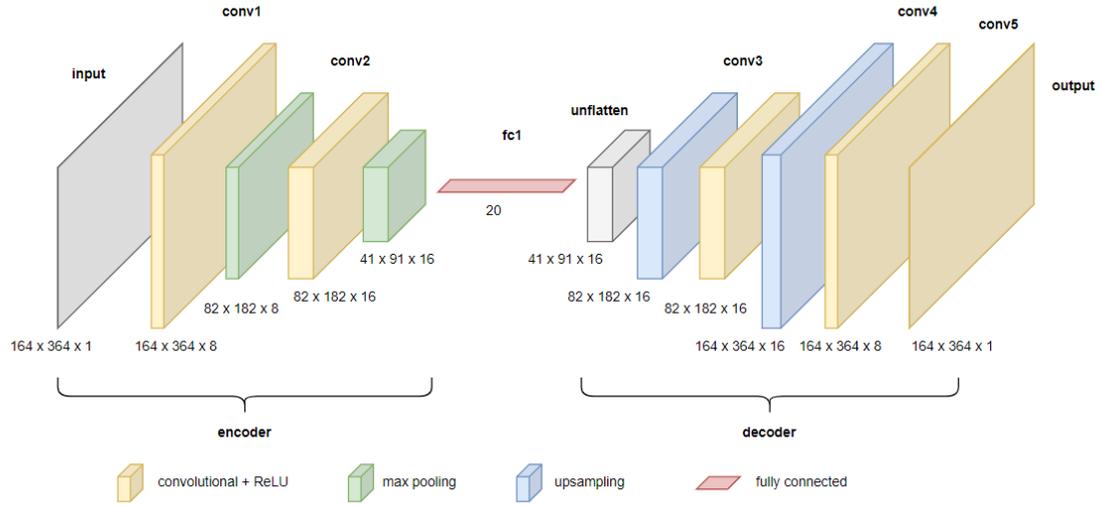

**Fig 2:** Autoencoder network architecture

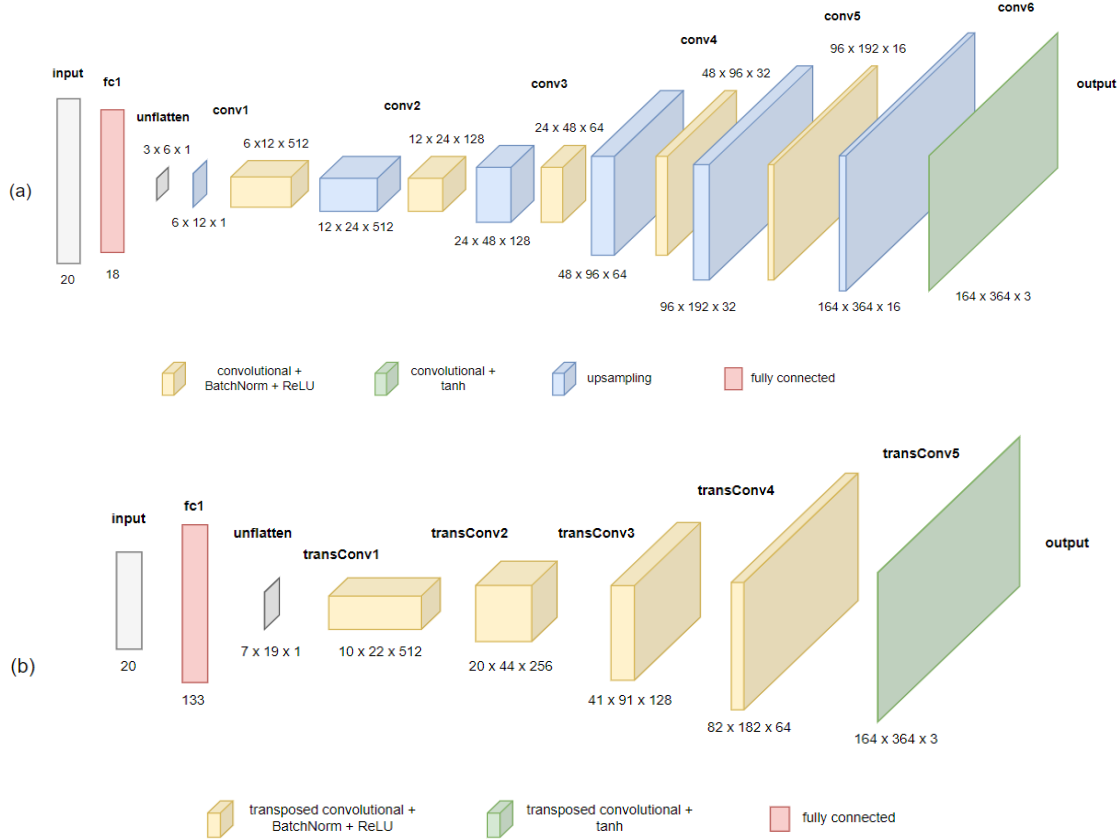

**Fig. 3** (a) upsample-Generator and (b) transConv-Generator network architectures

In the context of CFD simulations, cGANs have shown promise in generating more accurate and high-fidelity results by conditioning the generator on relevant simulation parameters such as Reynolds number, angle of attack, and turbulence intensity. This allows the generator to produce pressure distribution plots tailored to specific flow conditions and geometries. By incorporating such conditional information into the training process, cGANs can learn to generate more realistic and physically meaningful CFD results, significantly reducing the computational cost and time required for traditional CFD simulations.

In a cGAN architecture, the generator $G$ takes random noise $z$ and the conditional information $y$ as input and generates synthetic samples. The discriminator $D$ takes both real samples $x$ and their corresponding conditional information $y$ as input and tries to distinguish between real and synthetic samples. Objective for the cGAN can be formulated as:

$$min_G max_D \mathcal{L}_{cGAN}(G,D) = \mathbb{E}_{x,y}[logD(x,y)] + \mathbb{E}_{y,z}\left[\log\left(1 - D\big(G(y,z)\big)\right)\right] \quad (4)$$

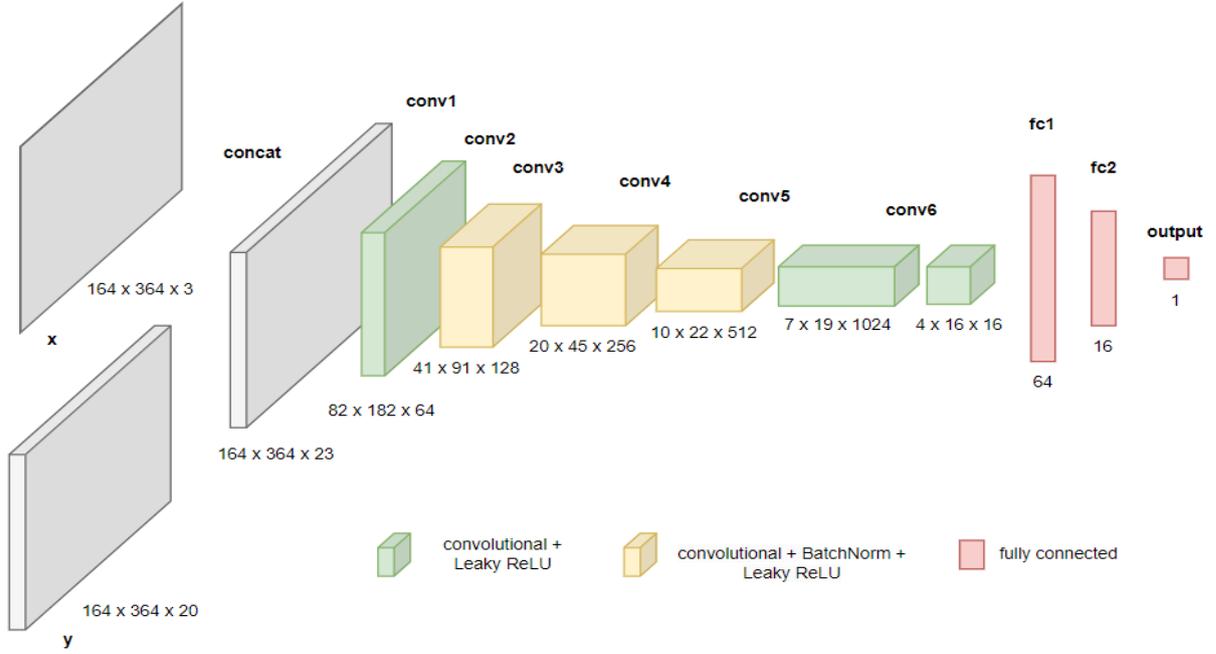

**Fig 4:** Discriminator network architecture

In our implementation, the GAN is formulated in a fashion similar to the cGAN, which is conditioned to generate an output based on a specific class. Instead of taking noise as an input as done traditionally, the Generator $G$ takes in the latent vector $y$ generated by the autoencoder as the input, which acts as the conditional information. This vector $y$ is also passed into the discriminator after resizing, and is concatenated with the input, which is either the real data $x$, or the generated data $G(y)$. The objective function for this GAN formulation can be expressed as:

$$min_G max_D \mathcal{L}_{GAN}(G, D) = \mathbb{E}_{x,y}[logD(x,y)] + \mathbb{E}_y\left[\log\left(1 - D(G(y))\right)\right] \quad (5)$$

An L1 loss term between the generated image and the expected image output is added to the Generator loss, which measures the average absolute pixel-wise difference between the corresponding pixels of the two images. The L1 loss encourages the generator to produce outputs that closely match the target image in pixel intensities. This loss term can be expressed as:

$$\mathcal{L}_{L1}(G) = \mathbb{E}_{x,y}[\|x - G(y)\|_1] \quad (6)$$

The final objective now becomes:

$$\mathcal{L}(G, D) = min_G max_D \mathcal{L}_{GAN}(G, D) + \lambda \mathcal{L}_{L1}(G) \quad (7)$$

where $\lambda$ is a hyperparameter denoting the weight of the L1 loss. Two variants of the generator are proposed: one with upsampling layers (upsample-Generator) and the other with 2D transposed convolution layers (transConv-Generator).

The upsampler-generator uses five upsampling layers followed by 2D convolution layers, along with ReLu activations (except tanh after the last layer) and batch normalization. The transConvranspose-generator uses transposed convolutions to replace upsampling + 2D convolution layers. The two network architectures are given in Fig. 3.

The Discriminator $D$ consists of convolutional layers, batch normalization and LeakyReLU activation. It ends up with a single output with sigmoid activation, which gives a probability that classifies the given input as either real or fake. The discriminator is constructed keeping the guidelines given by Radford et al. [20] in mind. Strided convolution, as opposed to pooling, is recommended for downsampling because it allows the network to develop its own pooling function. Additionally, healthy gradient flow is promoted by batch norm and leaky ReLU functions, which are essential for both the Generator $G$ and the Discriminator $D$'s learning processes. Fig. 4 shows the Discriminator network architecture.

During training, a sample first passes through the autoencoder, which extracts its important features in the form of a 20-dimensional vector, and this vector can now be used as an input to the GAN. The Generator $G$ takes in this vector and generates an output which is supposed to be a mapping to the input's pressure coefficient field. Such fake samples, along with the true samples are passed to the Discriminator $D$, with their corresponding extracted feature vectors reshaped to the shape [batch_size, 164, 364, 20]. This shape is the same height and width as the field but with 20 channels instead of 3. Inside the network, these are concatenated and passed through the subsequent layers to give a classification result finally. As formulated before, the generator loss combines the usual adversarial loss and the L1 reconstruction loss between the true and the generated image, as formulated before. The hyperparameter λ is the weight applied to the L1 reconstruction loss term, and is chosen as 500 after tuning.

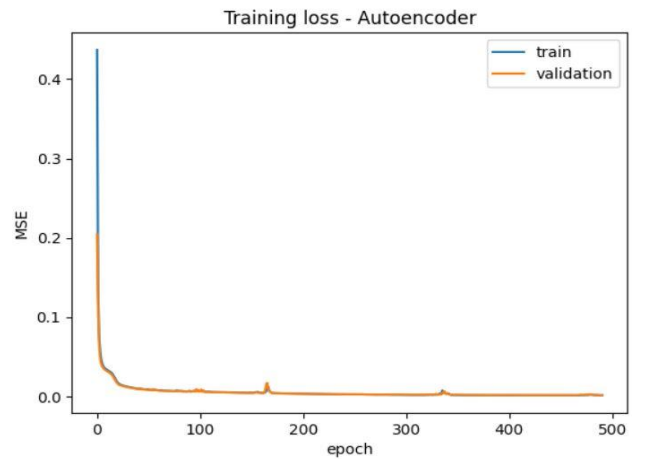

**Fig. 5** Autoencoder loss

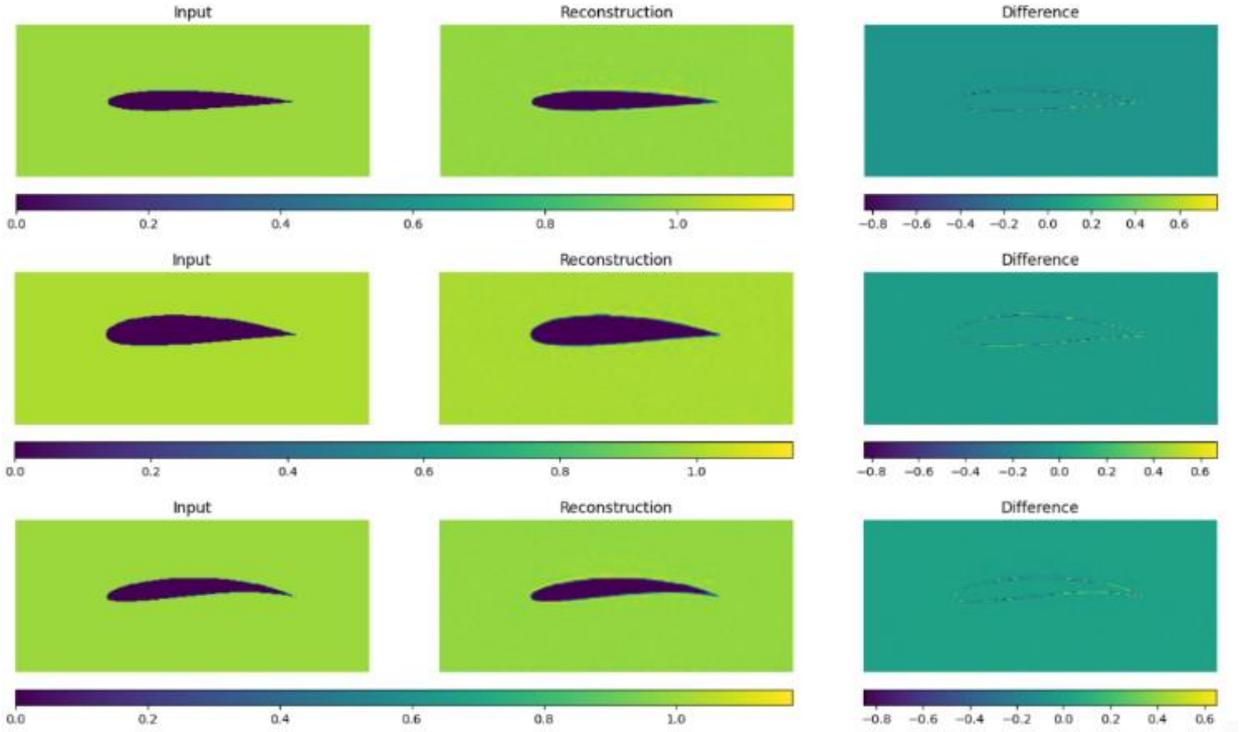

**Fig. 6** Autoencoder reconstructions on some test samples

### III. RESULTS AND DISCUSSION

*A. Autoencoder*

The autoencoder, consisting of 2D convolutional layers, max pooling, dense layers, and ReLU activations, learns to reconstruct the input geometry of shape [164x364x1], passing it through a bottleneck. A latent vector of size 20 seemed sufficiently capable of reconstructing the inputs. This latent vector attempts to capture the essential features of the geometry represented by the input to reconstruct the output based on this compressed feature representation. The training was done using 4096 samples, of which 3276 were used as training samples and 820 as validation samples. The network weights were updated using the Adam optimizer, which computes individual adaptive learning rates for different parameters from estimates of the first and second moments of the gradients. A learning rate of 1e-3 and a batch size of 16 were used. The MSE loss between the original input and the reconstruction is minimized, during which a meaningful latent representation of the input is derived in the bottleneck.

The Autoencoder model is trained for 500 epochs, after which the training and validation losses converge. Fig. 5 shows the Autoencoder loss while training. The results of the reconstructions on some test samples are shown in Fig. 6. The reconstructions agree with the respective inputs fairly; however, a slight difference is observable along the boundary of the aerofoil. Despite this, the model has generalized well

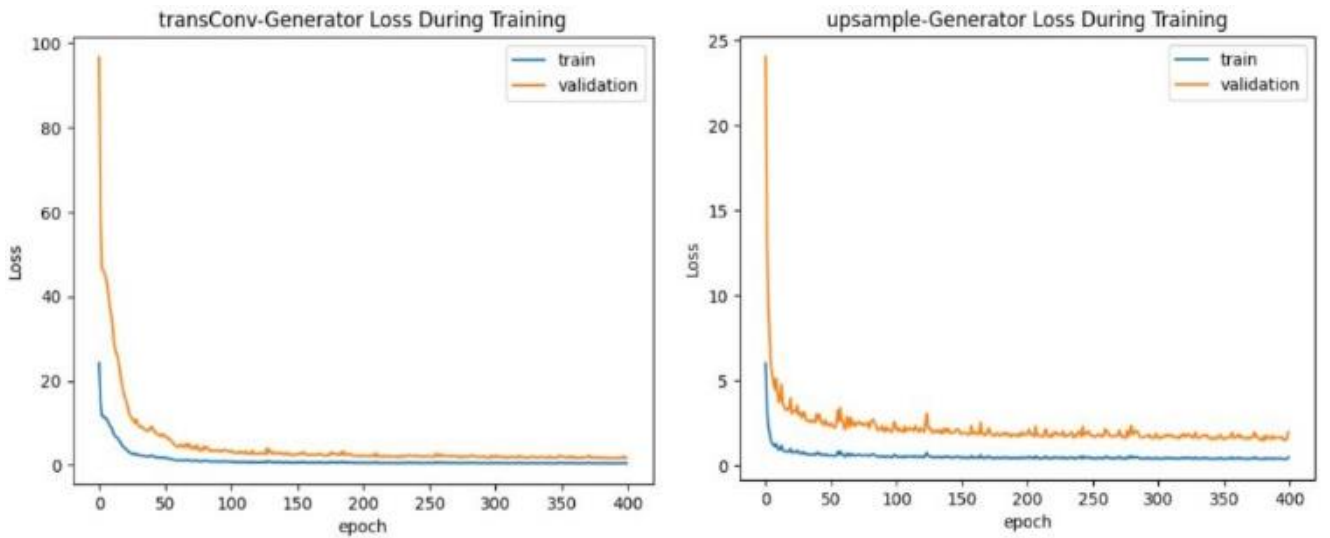

**Fig. 7** (a) transConv-Generator loss and (b) Upsample-Generator loss during training

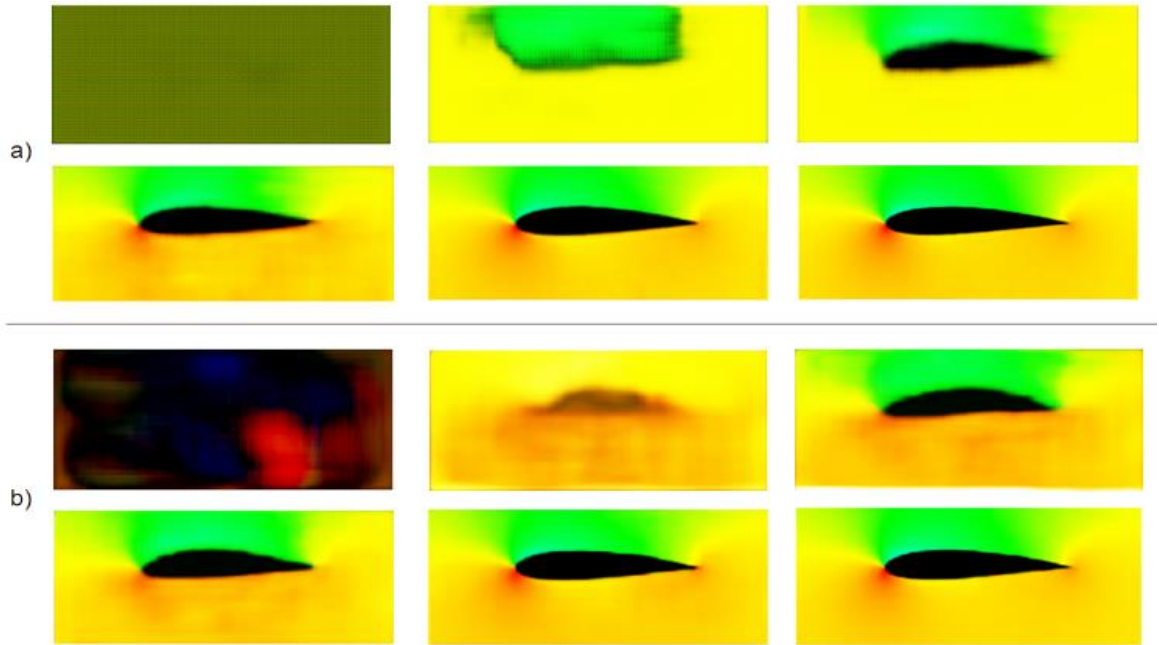

**Fig. 8** (a) transConv-Generator and (b) Upsample-Generator generating the field for a test sample for different aerofoil thicknesses, camber, and camber locations. Hence the encoder can now be detached and used as an independent compressor, which can convert a [164x364x1] raw input to its compressed representation of size [20].

*B. Conditional GAN*

Two variants of the GAN are trained, consisting of batch normalization, dense layers and ReLU activations, along with upsample + 2D convolutional layers or 2D transposed connvolutional layers. The Generator $G$, conditioned on the input latent vector, learns to generate pressure field corresponding to the input. At the same time the discriminator $D$ aims to classify between real and fake images, again looking at the expanded latent vector representation of the aerofoil along with the pressure field. In both the networks, the weights were updated using the Adam optimizer, and a learning rate of 1e-3 was used. The dataset was passed in with a batch size of 16.

The networks are trained together for 400 epochs, by the end of which the losses converge. Fig. 7 shows the Generator losses for the two variants while training. The generation of a validation sample as the model trains is shown in Fig. 8. In the initial epochs, the transConv-Generator exhibits artifacts which are generally caused by "uneven overlap" of the convolution kernel [19] but they diminish by the end of 400 epochs. Overall by the end of 400 epochs, both the generators achieve similar results. However, the aerofoil surface in the case of transConv-Generator looks smoother as compared to the upsample-Generator.

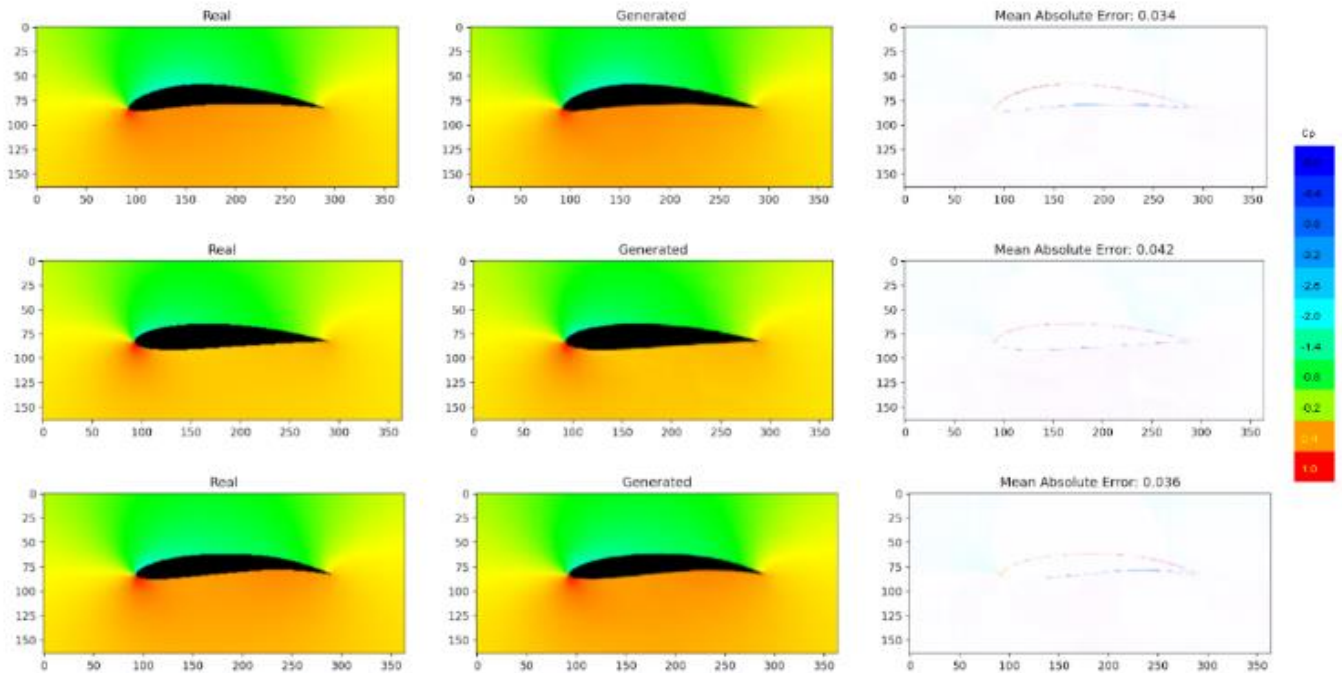

**Fig. 9** Real and Generated test samples using upsample-Generator

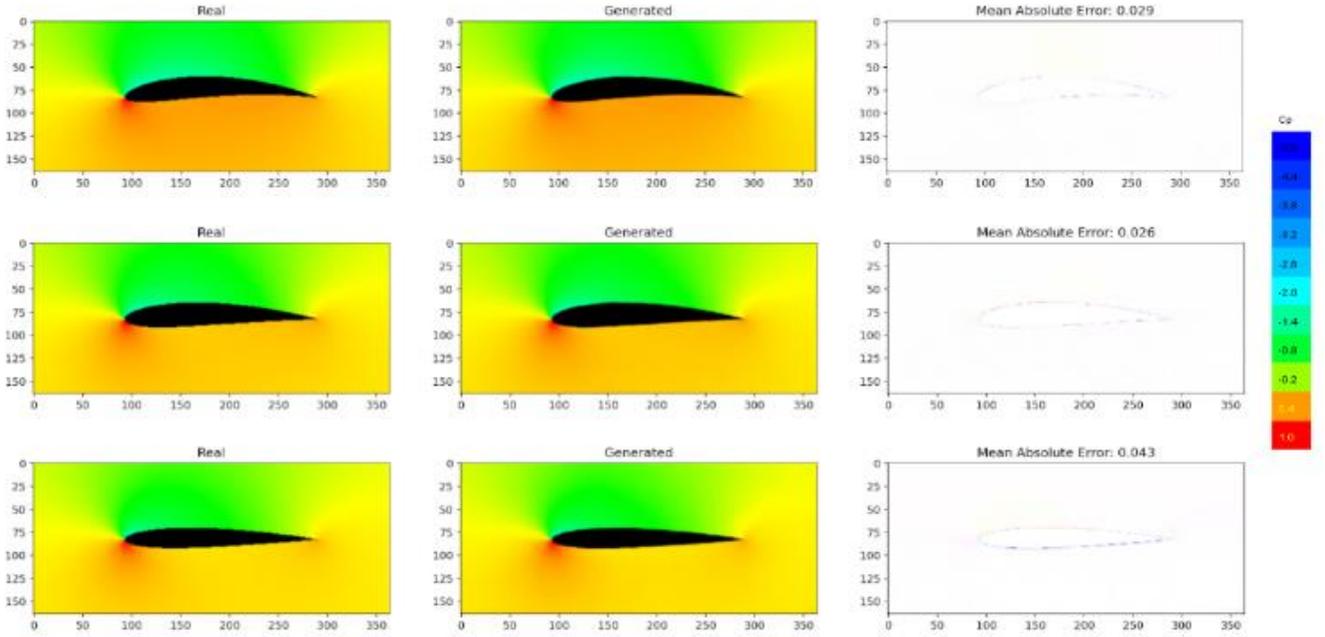

**Fig. 10** Real and Generated test samples using transConv-Generator

*C. Pressure field prediction*

The encoder and generator weights saved after training completion are used to test the results of the combined model. The input geometry of size [164x364x1] is passed into the encoder, which outputs a vector of size 20. This vector is then resized and passed into the Generator G which generates an output of size [164x364x3], conditioned on the input vector. The predictions using the upsample-Generator are shown in Fig. 9, and using the transConv-Generator are shown in Fig. 10. It can be seen that there is a good match between the real and the generated pressure fields with sufficiently low mean absolute error (MAE) values. Higher values of error are concentrated along the aerofoil boundary, and seem less pronounced in the regions with lower $C_p$ values. It is evident from the results that the combined AE+GAN model can generate predictions with a high degree of accuracy over varying geometry parameters like thickness, camber, and camber location, accommodating various aerofoil profiles.

The performance of the two generator variants on a set of test samples is compared in figure 11. It can be observed that the MAE values in case of the transConv-Generator are almost always lower than when upsample-Generator is used. This is likely due to the fact that the transposed convolutional layers, being trainable, can learn the spatial upsampling. MAE for the value of the pressure coefficient ranges for the transConv-Generator ranges from 0.025 to 0.043, while for the upsample-Generator lies between 0.031 and 0.052. Overall, the real and predicted fields are in good agreement, with consistent predictions around varying aerofoil geometry.

## IV. CONCLUSIONS

The study results demonstrate that the combined AE+GAN model successfully generates predictions with high accuracy. The generated pressure fields closely match the real pressure fields, as indicated by the low MAE values. The errors observed primarily occur along the aerofoil boundary and are less pronounced in regions with lower Cp values. Nevertheless, the model accommodates various aerofoil profiles, including varying geometry parameters such as thickness, camber, and camber location.

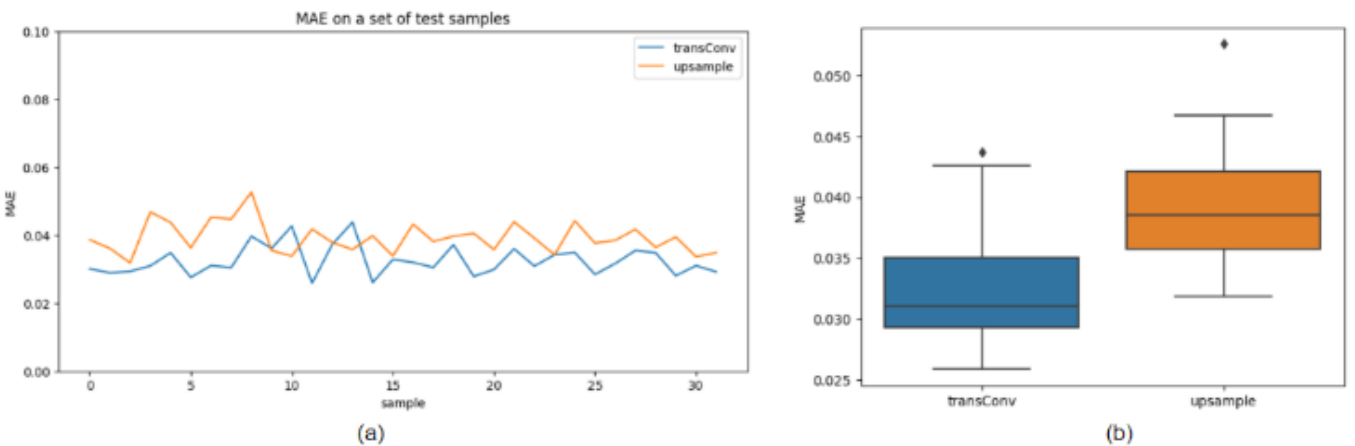

**Fig. 11** Comparison of MAE for the two Generator variants

This research contributes significantly to the advancement of computational techniques in fluid dynamics, specifically in the domain of aerodynamics. By leveraging the synergistic power of autoencoders and GANs, the proposed approach demonstrates the potential to reduce the time and costs associated with aerodynamic prediction. The efficient evaluation of aerofoil performance facilitated by this approach opens up new avenues for enhanced design and optimization processes in aerodynamics.

Further refinements and investigations can be conducted to enhance the model's performance and address the observed errors along the aerofoil boundary. Additionally, exploring the scalability and applicability of this approach to more complex aerodynamic scenarios and real-world applications could be a valuable direction for future research.

## V. REFERENCES

[1] LingJ., Kurzawski, A., & Templeton, J. (2016). Reynolds averaged turbulence modelling using deep neural networks with embedded invariance. Journal of Fluid Mechanics, 807, 155-166.

[2] Kutz, J. N., Brunton, S. L., Brunton, B. W., & Proctor, J. L. (2017). Dynamic mode decomposition-based reduced-order modelling of complex fluid flows. Proceedings of the Royal Society A: Mathematical, Physical and Engineering Sciences, 473(2201), 20160499.

[3] Wang, Q., Zhang, X., Xie, Y., & Zhang, X. (2019). Optimization of a microfluidic mixer using genetic algorithm and machine learning. Computers & Chemical Engineering, 126, 218-227.

[4] Lapeyre, C. J., Hickey, J. E., & Brenner, M. P. (2018). Artificial neural networks trained on DNS data for turbulent flow reconstruction and prediction. Journal of Computational Physics, 366, 29-48.

[5] Jin, J. M., Zhang, W. W., Liu, L., & Shao, X. M. (2018). A generative adversarial network for generating synthetic data of unsteady turbulent flows. Journal of Computational Physics, 366, 1-15.

[6] Wang, Q., & Ling, J. (2020). Machine learning of accurate turbulence models from DNS data. Journal of Computational Physics, 404, 109105.

[7] Smith, R., Zhu, Y., & Eldred, M. S. (2017). Machine learning for UAV aerodynamic performance prediction. Journal of Aircraft, 54(2), 712-725.

[8] Zhang, M., Li, J., & Li, Z. (2018). Flow pattern recognition in multiphase flows using deep learning. Computers & Fluids, 173, 189-199.

[9] Chen, Z., Hu, X., Huang, P., & Wang, L. (2019). Recurrent neural networks for turbulence modelling. Journal of Fluid Mechanics, 878, 708-731.

[10] Liu, L., Zhang, W. W., Jin, J. M., & Ling, J. (2020). Data assimilation with machine learning for weather forecasting. Journal of Computational Physics, 408, 109287.

[11] Guo, Z., Shi, X., Zhang, L., Zhao, L., & Jiang, Z. (2021). Deep reinforcement learning for aerodynamic shape optimization. Aerospace Science and Technology, 111, 106546.

[12] Li, H., Ling, J., & Perdikaris, P. (2022). Transfer learning accelerated computational fluid dynamics. Journal of Computational Physics, 455, 110965.

[13] Wang, X., Zhang, J., Cao, W., & Kang, Q. J. (2022). Hybridizing physics-based models with machine learning for multiphase flow simulations. Journal of Computational Physics, 452, 110993.

[14] Wang, Q., Perdikaris, P., Karniadakis, G. E., & Koumoutsakos, P. (2018). A deep learning approach for turbulence modelling in complex flows. Physics of Fluids, 30(5), 055109.

[15] S. Chen, B. Mulgrew, and P. M. Grant, "A clustering technique for digital communications channel equalization using radial basis function networks," IEEE Trans. on Neural Networks, vol. 4, pp. 570-578, July 1993.

[16] J. U. Duncombe, "Infrared navigation—Part I: An assessment of feasibility," IEEE Trans. Electron Devices, vol. ED-11, pp. 34-39, Jan. 1959.

[17] C. Y. Lin, M. Wu, J. A. Bloom, I. J. Cox, and M. Miller, "Rotation, scale, and translation resilient public watermarking for images," IEEE Trans. Image Process., vol. 10, no. 5, pp. 767-782, May 2001.

[18] Vincent, P., Larochelle, H., Bengio, Y., & Manzagol, P. A. (2008, July). Extracting and composing robust features with denoising autoencoders. In *Proceedings of the 25th international conference on Machine learning* (pp. 1096-1103).

[19] Kingma, D. P., & Welling, M. (2013). Auto-encoding variational bayes. *arXiv preprint arXiv:1312.6114*.

[20] Radford, A., Metz, L., & Chintala, S. (2015). Unsupervised representation learning with deep convolutional generative adversarial networks. *arXiv preprint arXiv:1511.06434*.]

[21] Odena, A., Dumoulin, V., & Olah, C. (2016). Deconvolution and checkerboard artifacts. Distill, 1(10), e3.